\documentclass{ws-ijmpc}

\usepackage[ansinew]{inputenc}
\usepackage{amsmath} 
\usepackage{amssymb}
\usepackage{latexsym}
\usepackage{lscape}

\newcommand{\ud}{\mathrm{d}}

\begin{document}
\title{Emergence of multiscaling in heterogeneous complex networks}
\author{A. Santiago and R. M. Benito}
\address{Grupo de Sistemas Complejos,\\
   Departamento de F\'\i sica y Mec\'anica,\\
   Escuela T\'ecnica Superior de Ingenieros Agr\'onomos,\\
   Universidad Polit\'ecnica de Madrid,\\
   28040 Madrid, Spain.\\
   antonio.santiago@upm.es, rosamaria.benito@upm.es}
\date{\today}

\maketitle

\abstract{In this paper we provide numerical evidence of the richer behavior of the connectivity degrees in heterogeneous preferential attachment networks in comparison to their homogeneous counterparts. We analyze the degree distribution in the threshold model, a preferential attachment model where the affinity between node states biases the attachment probabilities of links. We show that the degree densities exhibit a power-law multiscaling which points to a signature of heterogeneity in preferential attachment networks. This translates into a power-law scaling in the degree distribution, whose exponent depends on the specific form of heterogeneity in the attachment mechanism.}
\smallskip \par \noindent 
$Keywords:$ complex networks, power laws, scaling, multiscaling, preferential attachment, heterogeneous networks.
\smallskip \par \noindent
$Pac \,\, numbers:$ 89.75.Fb, 89.75.Hc, 89.75.-k.

% Texto

\section{Introduction}

A complex network is a set of nodes and links with a non-trivial topology (without evident principles of design) \cite{str01}. Networks pervade all domains of science: examples can be found in the natural \cite{wuc03}, social \cite{wu03}, technological \cite{yoo02} and cultural \cite{fer01a} realms. In the current effort to achieve a single coherent framework for complex systems, network theory has focused on the underlying principles that govern their topology. The coupling architecture of a system is an important ingredient in its modeling because its topology affects its collective dynamics and thus its function. Through a combination of empirical observations, analytical insight and numerical simulation, complex networks have been subject to intensive scrutiny during the last years \cite{alb02,new03a}. 

Dynamical network models \cite{dor02a} are discrete-time (generally stochastic) dynamical systems that evolve networks by the iterated addition/subtraction of nodes/links. These regard network topology as an emergent property of its evolution, focusing on the mechanisms that concur on such process. Among these mechanisms, preferential attachment enjoys a leading position in network literature. It was first discovered by Simon \cite{sim55} in the 1950s and applied to citation networks by Price \cite{pri65,pri76} in the 1970s. Recently, the so-called preferential attachment model by Barabási and Albert \cite{bar99a} has adopted this mechanism for the generation of undirected networks. The preferential attachment model provides a minimal account of sufficient mechanisms for the emergence of \emph{scale-free} networks \cite{bar03a}. Such networks are characterized by a connectivity distribution according to a power law, $P(k) = k^{-\gamma}$, which leads to a non-negligible presence of \emph{hubs} (highly connected nodes). 

In this paper we present numerical results concerning the connectivity degrees in the threshold model. The threshold model belongs to a general class of heterogeneous preferential attachment models where the affinity between node states biases the connectivity degrees in the attachment rule. First we review the basic facts about the preferential attachment model. Next we introduce our framework for heterogeneous preferential attachment models, and we define the threshold model. Afterwards we briefly discuss the methodology followed in the numerical simulation and we discuss the results concerning the degree distribution and degree densities of the model, comparing them with the analytical predictions made in the thermodynamic limit. We end this paper with concluding remarks about the significance and possible applications of the results.

\section{Preferential attachment}

The preferential attachment model by Barabási and Albert \cite{bar99a} is a dynamical network model that prescribes two mechanisms in the evolution of a network: growth and preferential attachment. The process starts with a seed of arbitrary size, and a new node is added to the network at each step. Each newly added node has a fixed number $m$ of links attached, which are connected to the already existing nodes following the so-called \emph{attachment rule}. The rule states that the linking probability of a network node $v_i$ is proportional to its connectivity degree $k_i$, $\Pi (v_i) = k_i / \sum_j k_j$. The addition of nodes is iterated until a network with a desired size $N$ is achieved.

The properties of the preferential attachment model have been widely studied, using both analytical and numerical methods \cite{bar99b,dor00b,kra01}. The networks generated exhibit power law degree distributions with $\gamma \simeq 3$, which becomes $\gamma = 3$ in the thermodynamic limit $N \to \infty$ \cite{dor00b}, moderate clustering levels and short geodesic distances. The preferential attachment model is strictly topological as the node degrees are the only metric that drives network evolution. The model has been subject to a series of variations \cite{alb00,dor00,erg02} that are still \emph{homogeneous} in nature since the attachment rule doesn't take into account intrinsic properties of the nodes. Although network theory has led to a significant improvement in our understanding of complex systems, it has been argued that an augmentation of its framework is required in order to improve our modeling of complexity \cite{str01,ama04}.

More recently, heterogeneous networks have been proposed as a first step in addressing the complication introduced by the influence of individual elements on the network structure. \emph{Weighted} networks have been proposed \cite{yoo01,barr04,bart04} to account for differences in the strength of the connecting links. Other proposed models include social network models \cite{bog04,gra06} which make use of the so-called ``social distance''; a competition model \cite{bia01} which incorporates a fitness measure; a local knowledge model \cite{gom04} which spatially limits the attachment of links; a metric model \cite{soa05} which make the attachment probability proportional to a power of the Euclidean distance between nodes; a gas-like model \cite{thu05} which interprets nodes as molecules that can exchange links upon collision; and topological automata \cite{alo06} where both node states and links destinations depend on the states of neighboring nodes in a previous iteration .

\section{Heterogeneous preferential attachment}

In order to enable a systematic analysis of the influence of heterogeneity in preferential attachment networks we have introduced an extended formalism for heterogeneous models. The preferential attachment model can easily be generalized to heterogeneous networks by imposing a metric structure on the node states while preserving the original mechanisms of growth and preferential attachment. This provides an intermediate abstraction level between homogeneous networks and topological automata.

\subsection{Heterogeneous models with global affinity}

In this section we formally define a general class of heterogeneous models where node states bias the attachment probabilities of links in a generalized attachment rule.
\begin{definition}~\\
A heterogeneous preferential attachment model with global affinity $M_1$ is a $3$-tuple $(R,\rho,\sigma)$, where:\\
(1) Each network node $v_i$ is characterized by a \emph{state} or \emph{attribute} $x_i$ on an arbitrary metric space $R$.\\
(2) Each node state $x_i$ is randomly assigned following a distribution $\rho$ over the state space $R$.\\
(3) Each pair of nodes $v_i$ and $v_j$ has a nonnegative affinity $\sigma$ depending on the two state variables $x_i$ and $x_j$.
\end{definition}
The class of heterogeneous preferential attachment models with global affinity $HPA_g$ is the set of all $3$-tuples that satisfy the conditions in Definition 1: $HPA_g = \{M_i : M_i = (R,\rho,\sigma)\}$. The formalism introduced defines the evolution of a network according to the following rules:

\noindent (i) The node states describe intrinsic properties which are deemed constant in the timescale of evolution of the network. The network links $e_i$ are not characterized by any state or affinity.

\noindent (ii) The growth process starts with a seed composed by $N_0$ nodes and $L_0$ links. The seed nodes $v_i$ are assigned arbitrary states $x_i \in R$.

\noindent (iii) A new node $v_a$ is added to the network at each iteration of the process. The newly added node $v_a$ has $m$ links attached to it. The number $m$ is common for all the added nodes and remains constant during the evolution of the network. The newly added node is randomly assigned a state $x_a$ following the distribution $\rho(x)$ over $R$.

\noindent (iv) The $m$ potential links attached to the newly added node $v_a$ are randomly connected to the network nodes following a probability distribution $\{\Pi(v_i)\}$ given by a \emph{generalized attachment rule},
\begin{equation}
\label{eq:1}
\Pi (v_i) = \frac{\pi(v_i)}{\sum_j \pi(v_j)}.
\end{equation}
The visibility $\pi$ of a node $v_i$ in the attachment rule is given by the product of its connectivity degree $k_i$ and its affinity $\sigma$ with the newly added node $v_a$,
\begin{equation}
\label{eq:2}
\pi (v_i) = k_i \cdot \sigma(x_i, x_a).
\end{equation}
The function $\sigma$ measures the affinity of the interaction between nodes $v_i$ and $v_a$ as a function of their states $x_i$ and $x_a$. It thus can be seen in Eq. \ref{eq:2} that the affinity $\sigma$ for each potential interaction biases the connectivity degree $k_i$ of the candidate node in the attachment rule according to the states of the two nodes. The higher the affinity of a candidate node with the added node, the higher its probability $\Pi (v_i)$ of acquiring a potential link. Steps (iii) and (iv) are iterated until a desired number of nodes has been added to the network. Obviously, only nodes with a strictly positive visibility can be regarded as receptive to the links attached to a newly added node.

\subsection{Threshold models}

The choice of the tuple $(R, \rho, \sigma)$ in the previous section determines the form of heterogeneity in the attachment mechanism and thus the network topology. In order to apply numerical simulation we need to specify particular models within this general class. Next we define a family of heterogeneous preferential attachment models we refer as the threshold models. The threshold models are based on the assumption that the affinity between network nodes is proportional to the similarity of their intrinsic properties. In other words, the affinity $\sigma$ is inversely related to the distance between the node states as defined by the metric $d$ on the state space $R$. In the spirit of the original preferential attachment model, the threshold models provide a minimal account of jointly sufficient mechanisms for the emergence of scale-free heterogeneous networks.

\begin{figure}
\centering \includegraphics[width=0.85\textwidth]{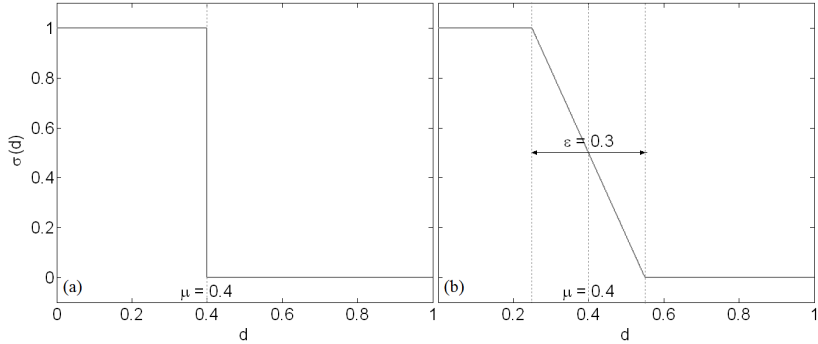}
\caption{Affinity $\sigma$ in the threshold model (Eq. \ref{eq:3}) as a function of the distance $d$ between states for a threshold $\mu = 0.4$ and different width $\epsilon$ values. The width values are: (a) $\epsilon = 0$, (b) $\epsilon = 0.3$.}
\label{fig:1}
\end{figure}

As a first approximation we will consider the case where the transition of $\sigma$ is discontinuous, this can be accounted by introducing a single parameter $\mu$ that defines the threshold distance for the transition. Then the affinity is a reverse step function of the distance between the node states and can be regarded as a form of local attachment \cite{gom04}, adopting the expression:
\begin{equation}
\label{eq:4}
\sigma (x_i, x_a) = 1 - H (d(x_i,x_a) - \mu) = 1 - H_{\mu} (d(x_i,x_a)), 
\end{equation}
where $H(x)$ is the Heaviside function. The \emph{interaction threshold} $\mu$ is shared by all the network nodes and governs the shape of the affinity in the preferential attachment rule, thus the form of heterogeneity in the network evolution. The shape of the affinity function $\sigma$ in this case is depicted in Fig. \ref{fig:1} (a) for a particular value of $\mu$.

Therefore, when the distance between the added node and the network node $d(x_i,x_a)$ is lower than the threshold $\mu$, the receiving node is regarded as a valid candidate for the attachment and its affinity is maximum, $\sigma = 1$. In such a case the visibility of the network node is given by its connectivity degree, $\pi (v_i) = k_i$. It should be noted that when the threshold $\mu$ is equal to the maximum possible distance, $d_{max}$, there is obviously a maximum affinity between all the network nodes and the added node, irrespective of their states. In such circumstances the effective heterogeneity in the attachment mechanism is null, the visibility of all the network nodes is equal to their connectivity degrees and the dynamics recover the behavior of the homogeneous preferential attachment model proposed by Barabási and Albert. 

On the other hand, when the distance $d(x_i,x_a)$ is larger than the threshold $\mu$ the affinity is inexistent, $\sigma = 0$. The lack of affinity between the nodes means that the visibility of the network node is zero from the standpoint of the newly added node, in this case their interaction is not possible since the attachment probability is $\Pi (v_i) = \pi (v_i) = 0$. Last, when the threshold $\mu$ is small enough so that a newly added node cannot find a candidate node in the whole network, then it is considered that this node is rejected by the network and it is not accounted by the network size $N$.

A more general case can be stated if we consider that the transition in the affinity $\sigma$ takes place smoothly along a transition regime. For simplicity, let us approximate the inverse relationship between affinity and distance by means of a piece-wise linear map. This can be accounted by introducing an additional parameter $\epsilon$ that defines the width of the transition regime. The shape of the affinity $\sigma$ in this second case is depicted in Fig. \ref{fig:1} (b). Thus we arrive at the following definition of a threshold model:

\begin{definition}~\\
A threshold model $TM_1 \in HPA_g$ is a $3$-tuple $(R,\rho,\sigma)$ where $\sigma : [0,d_{max}] \mapsto [0,1]$ is a nonnegative function defined for $0 < \mu \le d_{max}$, $0 \le \epsilon \le d_{max}$ as:
\begin{equation}
\label{eq:3}
\sigma (x_i, x_a) = \sigma (d(x_i, x_a)) = 
\left\{ \begin{array}{ll}
1 & \textrm{for } 0 \le d \le d_0 \\
1/2 + (\mu-d)/\epsilon & \textrm{for } d_0 < d < d_1 \\
0 & \textrm{for } d_1 \le d \le d_{max},
\end{array} \right.
\end{equation}
where $d$ is a metric on $R$, $d_{max} \equiv \mathrm{sup}_{x_1,x_2 \in R} \, d(x_1,x_2)$, $d_0 \equiv \max(0, \mu - \epsilon/2)$ and $d_1 \equiv \min(d_{max}, \mu + \epsilon/2)$.
\end{definition}

It should be noted that $0 \le \sigma \le 1$, hence the affinity between nodes takes the highest value when the distance $d \le d_0$ and the lowest value when $d \ge d_1$. The transition between these extreme affinity values takes place along a region defined by its middle point $\mu$ and its width $\epsilon$. Again, these two parameters are shared by all the network nodes and govern the shape of the affinity $\sigma$ in the preferential attachment rule. When the width $\epsilon = 0$ the affinity becomes a step function as considered in the previous case. When $\epsilon > 0$ the affinity $\sigma$ can take values along the interval $0 \le \sigma \le 1$ and the transition in the visibility of the nodes between $k_i$ and $0$ is smooth. Likewise, rejected nodes are not accounted by the network size $N$.

Henceforth we will focus on the unidimensional version of the threshold model over the real line, thus we will choose the unit interval $R = [0,1]$ as state space and the Euclidean distance $d = |x_i - x_a|$ as metric of the space state. The affinity parameters $\mu$ and $\epsilon$ of the model will therefore range in the interval $[0,1]$.

\section{Connectivity degrees in the threshold model}

Next we present the results concerning the behavior of the connectivity degrees in the threshold model over the real line as a function of the affinity parameters $\mu$ and $\epsilon$. The connectivity degrees in the networks have been characterized by means of the degree distribution and the degree densities. These metrics have been analyzed by numerical simulation of ensembles of networks, each of them characterized by a particular pair of affinity parameters. The other model parameters are shared by all the network ensembles. The distributions have been computed by means of histograms of the relevant metrics in the different ensembles of networks.

To implement the generalized attachment rule a vector of the visibilities of all incumbent nodes is calculated at each iteration according to Eq. \ref{eq:2}. A list of the nodes with positive visibility is then compiled, and a candidate node in this list is randomly chosen for each new link with a probability given by the normalized visibility according to Eq. \ref{eq:1}. Once a network node is selected and receives a new link at a given iteration, such node is removed from the list of candidate nodes during the remainder of the iteration. This follows from the fact that the network links are deemed as undirected and unweighted in the model, thus only one link can exist at most between a given pair of nodes.

The simulations have been carried out for different thresholds $\mu$ ranging in the interval $(0,1]$, and for each threshold several widths have been considered $\epsilon = \{0$, 0.1, $0.5\}$. The ensemble size is $10^4$ networks, a sample size chosen to render an accurate estimation of the distributions around the cutoff points induced by finite-size effects. Each network has a final size of $N = 10^4$ nodes, built from a random connected seed with $N_0 = 10$ nodes and $L_0 = 9$ links, through the addition of nodes with $m = 3$ links and states uniformly distributed over the unit interval. In order to evaluate the robustness of the behavior of the model, the analysis has been repeated for different network sizes $N$ and numbers of links $m$, without rendering qualitative discrepancies in the results.

\subsection{Degree distribution}

The degree distribution $P(k)$ of a network measures the probability for finding a node with degree $k$ in the network. Alternatively, the product $N \cdot P(k)$ measures the average number of nodes in the network with a given connectivity degree $k$. As it has been stated before, it should be emphasized that the main asset of the preferential attachment model lies in its ability to generate networks with a scale-free degree distribution, that is, asymptotically following a power law $P(k) \sim k^{-\gamma}$. Next we study how robust is such scaling against changes in the form of the affinity $\sigma$ of the threshold model, as defined by the parameters $\mu$ and $\epsilon$.

\begin{figure}[!t]
\centering \includegraphics[width=0.85\textwidth]{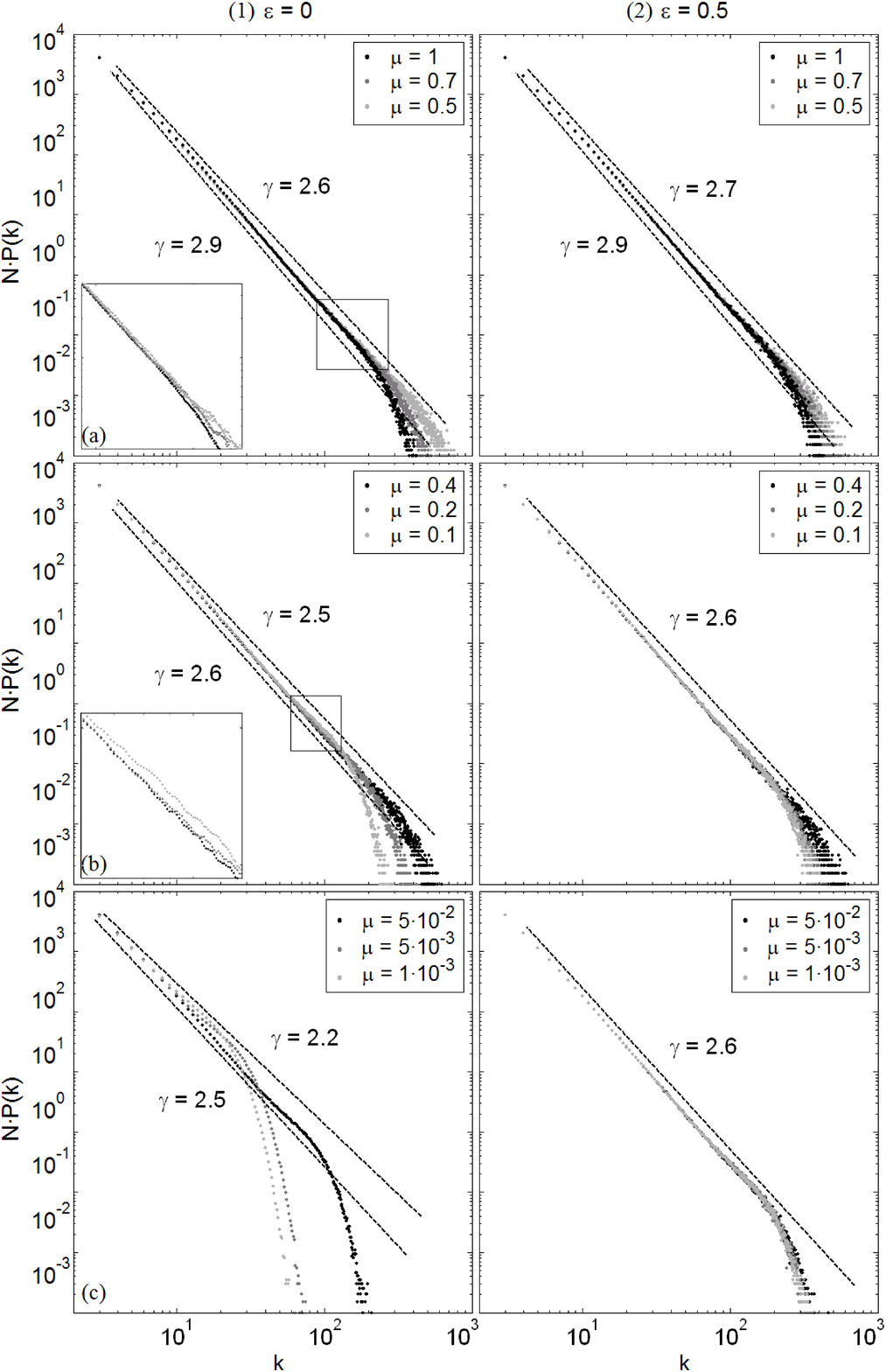}
\caption{Degree distribution $N \cdot P(k)$ in the threshold model for different values of the threshold $\mu$ and the width $\epsilon$. Rows (a), (b) and (c) correspond to $\mu = [0.5,1]$, $[0.05, 0.5]$ and $(0,0.05]$, respectively. Columns (1) and (2) correspond to $\epsilon = 0$ and $0.5$, respectively. Within each graphic, a grayscale differentiates the values of the threshold $\mu$. The other parameters are $\rho(x) \sim U(0,1)$, $S = 10^4$, $N = 10^4$, $N_0 = 10$, $L_0 = 9$ and $m = 3$.}
\label{fig:2}
\end{figure}

Fig. \ref{fig:2} shows the results of the numerical simulation for the degree distribution $P(k)$. Each curve depicted measures the average number of nodes $N \cdot P(k)$ with degree $k$ in a network generated with a particular vector of parameters. Alternatively, dividing the magnitude in the ordinate axis by the network size $N$ we obtain the average fraction of nodes with degree $k$ in a network. In the discussion of the results, let us focus first on the influence of $\mu$ when $\epsilon = 0$. This is the case when the affinity is a step function of the distance between the node states, $\sigma = 1 - H_{\mu} (|x_i - x_a|)$, and is represented in Fig. \ref{fig:2} Col. (1). Afterwards we will consider the effect of an increase in the transition width $\epsilon$. 

The curve $\mu = 1$, depicted in Fig. \ref{fig:2} (1.a), illustrates the homogeneous case ($\mu = 1$) equivalent to the original preferential attachment model. This curve asymptotically follows a power law $P(k) \sim k^{-\gamma}$ with an exponent $\gamma \simeq 2.9$. This value is equal to the exponent obtained by Barabási and Albert in numerical simulations of the preferential attachment model (see for instance Fig. 21 in Ref. \cite{alb02}). The slight discrepancy in the value of $\gamma$ with regards to the analytical prediction \cite{dor00b} is a finite-size effect that decreases the effective value of the exponent as the network size is reduced. Likewise, for larger degrees the distribution undergoes a crossover to an exponential decay, which is another finite-size effect due to the limited number of nodes in the simulated networks. The step-like plateaux are also a finite-size effect due to the relatively poorer statistics of hubs (that is, nodes with very high connectivity).

When the threshold value $\mu$ is modified, the numerical results in Fig. \ref{fig:2} (1.a)-(1.c) show that the power-law scaling in the degree distribution $P(k)$ is robust against changes in the heterogeneity of the attachment mechanism. The results also evidence that the variation in the scaling exponent $\gamma$ induced by changes in the form of heterogeneity is slight. These two traits exhibit a fair agreement of the numerical results with the behavior of the analytical solution for the stationary degree distribution of the model. In a previous paper \cite{san06} we showed that in the thermodynamic limit the degree distribution $P(k)$ of a heterogeneous model $M = (R,\rho,\sigma)$ in the $HPA_g$ class adopts for $k > m$ the expression:
\begin{equation}
\label{eq:5}
P(k) = \int_R \left( \prod_{j = m+1}^k \frac{\hat{w}(j-1)}{\hat{w}j+2} \right) \frac{2\rho}{\hat{w}m+2} \ud x 
\end{equation}
where $\hat{w}(x)$ is a normalized fitness factor,
\begin{equation}
\label{eq:6}
\hat{w}(x) \equiv \frac{w(x)}{\bar{w}} = \frac{w(x)}{\int_R w(x) \rho(x) \ud x}, \; \mathrm{with} \; w(x) \equiv \int_R \sigma(x,y) \rho(y) \ud y.
\end{equation}

\begin{figure}
\centering \includegraphics[width=0.80\textwidth]{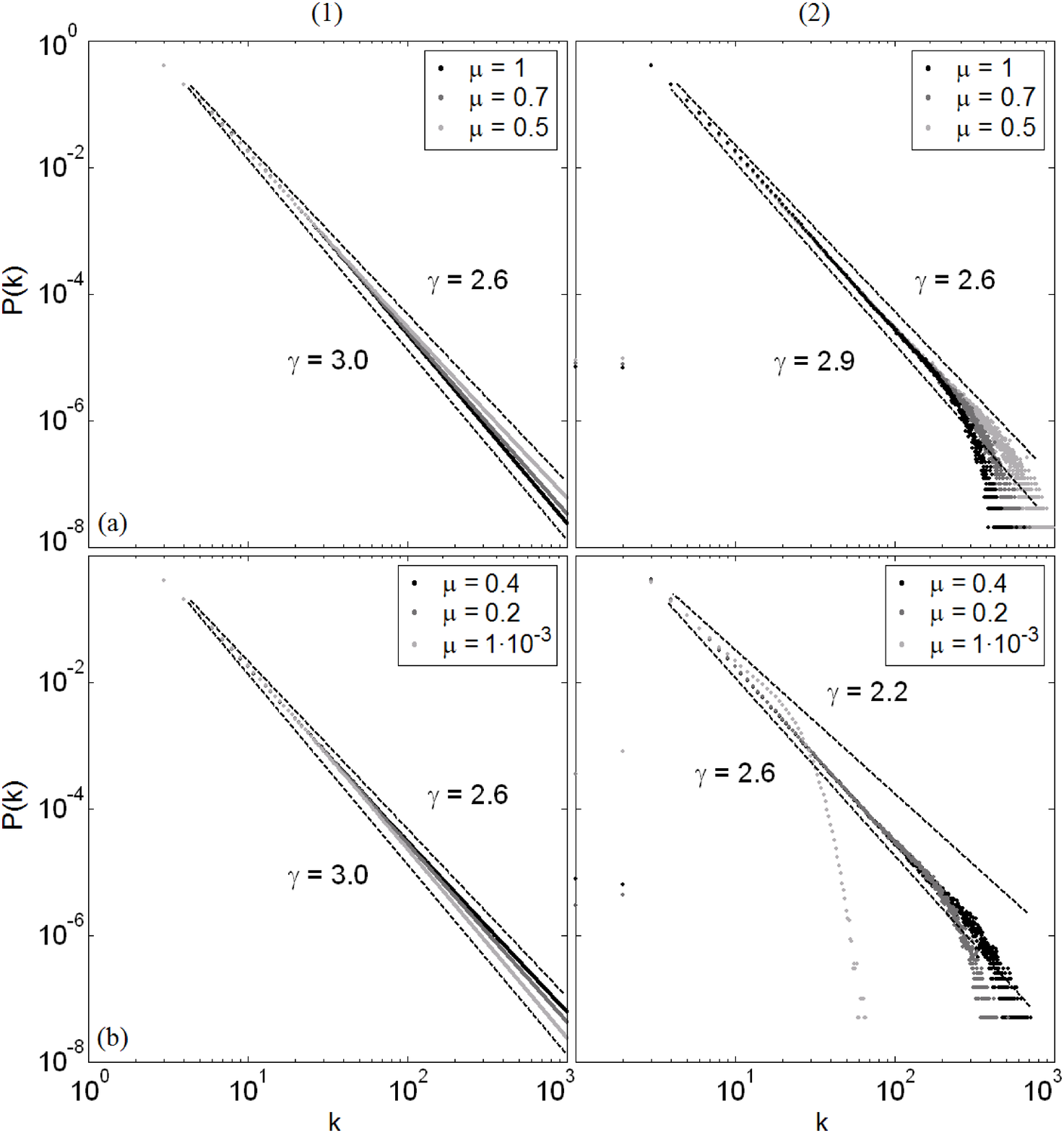}
\caption{Comparison of the analytical solution and the numerical simulation of the degree distribution $P(k)$ in the threshold model. Rows (a) and (b) correspond to $\mu = [0.5,1]$ and $(0,0.5]$, respectively. Column (1) corresponds to the analytical solution (Eq. \ref{eq:5}) with $\rho(x) \sim U(0,1)$, $\epsilon = 0$ and $m = 3$. Column (2) corresponds to the numerical estimation with $\rho(x) \sim U(0,1)$, $\epsilon = 0$, $S = 10^4$, $N = 10^4$, $N_0 = 10$, $L_0 = 9$ and $m = 3$.}
\label{fig:3}
\end{figure}

For a convenient comparison of the analytical and numerical results, Fig. \ref{fig:3} depicts the stationary degree distribution obtained from Eq. \ref{eq:5} for the threshold model, next to the degree distribution obtained by numerical simulation, when $\epsilon = 0$. A first evident trait in the numerical results shown in Fig. \ref{fig:2} (1.a) and (1.b) is the existence of a change of regime in the influence of $\mu$ observed around $\mu = 0.5$. When $\mu > 0.5$, Fig. \ref{fig:2} (1.a) evidences that a decrease in the threshold $\mu$ yields a lower scaling exponent ($\gamma \simeq 2.9$ for $\mu = 1$, while $\gamma \simeq 2.6$ for $\mu = 0.5$) and a higher cutoff degree in the distribution $P(k)$, which points towards a higher presence of hubs.

The behavior exhibited by the exponent $\gamma$ of the numerical distribution $P(k)$ as a function of a decreasing $\mu$ agrees with the behavior of the stationary distribution, which was analytically explained as a consequence of the introduction of density components $f(k,x)$ with a slower decay, which dominate the asymptotical behavior of $P(k)$. Alternatively, this behavior can be explained through the increasing inequality of the affinity regions of the network nodes over the state space $R$, which translates into an increase in the visibility of nodes with moderate states and a decrease in the visibility of nodes with peripheric states. As the threshold $\mu$ approaches 0.5, the differences in the visibility of the nodes grow, so that the network nodes with largest affinity regions face a lower competition in the acquisition of links and attain higher degrees.

It should be noted that as the threshold $\mu$ approaches 0.5, the exponent $\gamma$ in the analytical solution exhibits a slightly better agreement with the numerical results. This can be attributed to an interplay between finite-size effects and the inhomogeneity of the density components of $P(k)$. On the one hand, the decrease of the network size tends to magnify finite-size effects that yield a reduction in the exponent $\gamma$. On the other hand, the decrease of the network size tends to reduce the relative advange accrued by the most visible nodes, and thus yield an increase in the same exponent $\gamma$. Therefore, the inequality in the affinity regions brings a better agreement between the stationary and numerical distributions.

When the threshold $\mu < 0.5$, Fig. \ref{fig:2} (1.b) evidences that a decrease in $\mu$ still yields a lower exponent $\gamma$ ($\gamma \simeq 2.5$ for $\mu = 0.05$) as it was observed when $\mu > 0.5$. Furthermore, a decrease in $\mu$ also yields a reduction in the cutoff degree to the exponential regime, which points to a lower presence of hubs in the networks. The lower presence of hubs in the simulated networks is in agreement with the behavior of the stationary distribution, however the behavior of the exponent $\gamma$ shows a discrepancy with regards to the analytical predictions. The regime reversal exhibited by the presence of hubs in the networks was analytically explained as a consequence of the progressive removal of density components $f(k,x)$ with slower decay.

Alternatively, it should be noted that a reduction of $\mu$ below 0.5 yields a decrease in the inequality of the affinity regions of the network nodes over the state space $R$, which yields a weakening of the advantage accrued by the most connected nodes and a subsequent reversal in the cutoff degree. Concurrently, a decrease of $\mu$ also yields a growing localization of the attachment mechanism over the state space $R$ and a progressive degradation of the interaction domains of the newly added nodes. The fragmentation in the network dynamics translates into a magnification of the finite-size effects, a weakening of the positive feedback mechanism in the degree dynamics, and thus a further left shift in the cutoff degree. The growing finite-size effects also yield a reduction in the exponent $\gamma$, which dominates over the increase in $\gamma$ produced by the increasing equality in the balance of competition between the network nodes.

Finally, when the threshold $\mu \ll 1$, Fig. \ref{fig:2} (1.c) shows that the numerical distribution $P(k)$ continues the trend exhibited for larger thresholds in Fig. \ref{fig:2} (1.b). The scaling exponent $\gamma$ decreases slightly and the cutoff degree shifts to the left. In the limit when $\mu$ approaches zero, $P(k)$ converges to a power-law with an exponent $\gamma \simeq 2.2$. It should be emphasized that in the last case the finite-size effects introduce noticeable differences in the behavior of the numerical distribution against the stationary distribution, as evidenced by the lower scaling exponent and the reduced cutoff degree. In the thermodynamic limit $N \to \infty$, a reduction of the threshold from $\mu = 0.5$ to $\mu \to 0$ yields an increase in the scaling exponent from $\gamma \simeq 2.6$ to $\gamma \to 3$, as shown in Fig. \ref{fig:3} (1.b).

\begin{figure}
\centering \includegraphics[width=0.80\textwidth]{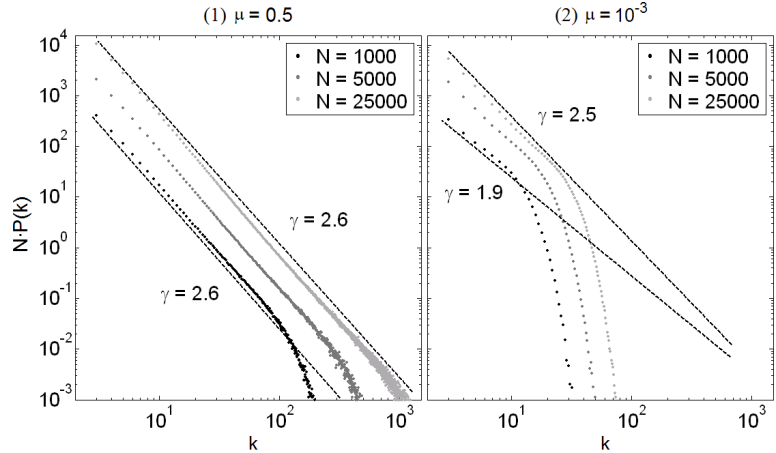}
\caption{Degree distribution $N \cdot P(k)$ in the threshold model for different network sizes $N$ and two values of the threshold $\mu$. Columns (1) and (2) correspond to $\mu = 0.5$ and $10^{-3}$, respectively. Within each graphic, a grayscale differentiates the network sizes $N$. The other parameters are $\epsilon = 0$, $\rho(x) \sim U(0,1)$, $S = 10^4$, $N_0 = 10$, $L_0 = 9$ and $m = 3$.}
\label{fig:3b}
\end{figure}

The variable influence of the finite-size effects on the degree distribution $P(k)$ as a function of the threshold value $\mu$ can be more clearly appreciated in Fig. \ref{fig:3b}, which illustrates the results of the numerical simulation for different network sizes and two threshold values. Clearly, larger network sizes yield higher cutoff degrees for the exponential regime irrespective of the threshold $\mu$ adopted. On the other hand, the influence of the network size $N$ on the scaling exponent of $P(k)$ largely depends on the threshold value $\mu$, which governs the degree of fragmentation of the preferential attachment on the state space. When $\mu$ takes moderate or high values, as in the case of Fig. \ref{fig:3b} (1), an increase in the size $N$ yields a slight increase in the exponent $\gamma$. As the threshold $\mu$ decreases and approaches zero, which is the case of Fig. \ref{fig:3b} (2), finite-size effects become more important and an increase in $N$ yields a progressively larger increase in the exponent $\gamma$.

To sum up, the growing inequality in the affinity regions of the network nodes is the main factor driving the observed phenomena when $\mu > 0.5$, while the dilution of the former inequality and the fragmentation of the attachment mechanism are the main factors driving the observed phenomena when $\mu < 0.5$. These factors alter the balance of competition for links in the preferential attachment mechanism and the subsequent presence of hubs in the networks. It should be emphasized that only the last factor has a finite-size origin, thus the agreement between the thermodynamic calculations and the numerical simulations (with reasonable network sizes) deteriorates as the heterogeneity level increases. These results prompts us to consider the heterogeneity in the attachment mechanism as an influential ingredient in the hub composition of real networks.

Next we scrutinize the influence of the transition width $\epsilon$ in the numerical degree distribution $P(k)$. When $\epsilon > 0$ the affinity is no longer a step function of the distance $d$ between node states, as illustrated in Fig. \ref{fig:1} (b). Despite the qualitative change in the form of $\sigma$, the simulation results furnish evidence that when $\epsilon$ is small enough (e.g. $\epsilon = 0.01$) the behavior of the distribution $P(k)$ is barely distinguishable from that obtained when $\epsilon = 0$, irrespective of the value of the threshold $\mu$ (results not shown). Therefore, the influence of the threshold $\mu$ on the degree distribution $P(k)$ is robust against the functional form of $\sigma$, in the sense that slight changes in $\epsilon$ do not introduce noticeable differences in the behavior of the distribution.

As the width $\epsilon$ takes higher values, the numerical results prove that the power-law scaling in the degree distribution $P(k)$ is preserved, as predicted by the analytical study. The results also show that as $\epsilon$ increases the sensitivity of $P(k)$ to changes in the threshold $\mu$ progressively decreases. When the threshold $\mu = 1$ the distribution is virtually the same for all widths $\epsilon$, however as $\mu$ is reduced from 1 to 0.5 the variation experienced by $P(k)$ is smaller as $\epsilon$ increases, as illustrated in Fig. \ref{fig:2} (2.a). As $\mu$ is reduced below 0.5 the variation experienced by $P(k)$ is even smaller for the lower half of the threshold interval, as illustrated in Fig. \ref{fig:2} (2.b). This is particularly evident for very small values of the threshold $\mu$, as represented in Fig. \ref{fig:2} (2.c).

The influence of the width $\epsilon$ on the stationary distribution $P(k)$ of the threshold model was analytically explained by the introduction of quadratic terms in the normalized fitness $\hat{w}$, which tend to dampen the dependence of the lowest scaling exponents $\gamma$ in the density components of $P(k)$ as a function of the threshold $\mu$ for higher $\epsilon$ values. Alternatively, this phenomenon may be explained by the shape of the affinity function $\sigma$ pictured in Fig. \ref{fig:1}. As the width $\epsilon$ increases, the transition regime in $\sigma$ becomes slower and this performs a sort of averaging on the interaction domains of the added nodes: more nodes are included in the domains, however the visibility of some is downgraded by the transition regime. This averaging effect dampens the dependence of the distribution $P(k)$ as a function of the threshold $\mu$, in particular for small thresholds $\mu$ when the shape of the $\sigma$ function becomes dominated by the transition regime.

\subsection{Degree density}

The degree density $f(k,x)$ of a heterogeneous network measures the probability density for finding a node with degree $k$ and state $x$ in the network. Assuming an uncountable set of states over the space $R$, as in the case of the threshold model studied in this paper, then $\int_{x_1}^{x_2} f(k,x) \ud x$ measures the probability for finding a node with degree $k$ and state $x_1 < x < x_2$. Our analytical studies of heterogeneous preferential attachment models \cite{san06} suggest that the presence of heterogeneity in the attachment mechanism leaves a signature in the topology of the networks that can be empirically checked. This signature is a power-law multiscaling in the degree densities according to exponents spanning a continuous interval around the homogeneous exponent $\gamma = 3$. The stationary density components $f(k,x)$ adopt the expression for $k \ge m$:
\begin{equation}
\label{eq:7}
f(k,x) = \frac{2\rho/\hat{w}}{m+2/\hat{w}} \frac{B(k, 1+2/\hat{w})}{B(m, 1+2/\hat{w})},
\end{equation}
where $B$ is the Legendre Beta function,
\begin{equation}
\label{eq:8}
B(y,z) = \int_0^1 t^{y-1} (1-t)^{z-1} \ud t \; \mathrm{for} \; y,z>0.
\end{equation}
This means that the asymptotic behavior of $f(k,x)$ when $k \to \infty$ is as
\begin{equation}
\label{eq:9}
f(k,x) \sim B(k,1 + 2/\hat{w}) \sim k^{-(1 + 2/\hat{w})}.
\end{equation}
so that the scaling exponents adopt the form:
\begin{equation}
\label{eq:10}
\gamma(x) = 1 + 2/\hat{w}(x).
\end{equation}

\begin{figure}
\centering \includegraphics[width=0.80\textwidth]{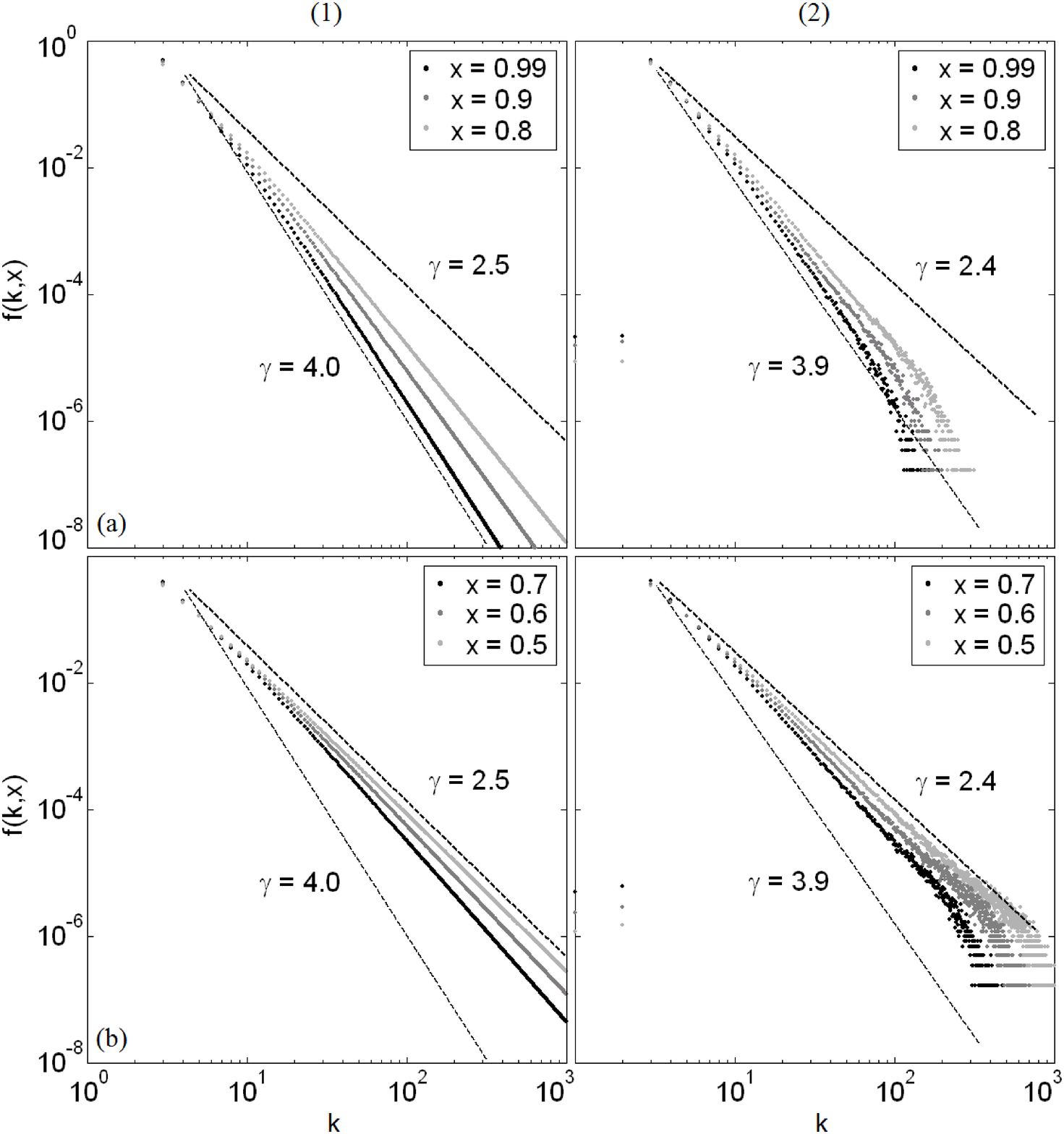}
\caption{Comparison of the analytical solution and the numerical simulation of the degree densities $f(k,x)$ in the threshold model for different node states $x$. Column (1) corresponds to the analytical solution (Eq. \ref{eq:9}) with $\rho(x) \sim U(0,1)$, $\mu = 0.5$, $\epsilon = 0$ and $m = 3$. Column (2) corresponds to the numerical approximation by means of the partial degree distributions $P(k|x)$ with $\rho(x) \sim U(0,1)$, $\mu = 0.5$, $\epsilon = 0$, $m = 3$ and $r_x = 0.01$. Within each graphic a grayscale differentiates the states $x$ that characterize each subset of network nodes. The other parameters are as in Fig. \ref{fig:3}.}
\label{fig:4}
\end{figure}

Next we present the results of the simulation of the degree densities that validate this prediction in the case of the threshold model and provide a numerical evidence of the existence of the multiscaling phenomenon. In order to compute the degree densities $f(k,x)$ we have estimated partial degree distributions $P(k|x)$ over subsets of network nodes with states in an interval on certain $x_0$ values of width $r_x$, $x \in (x_0 - r_x, x_0 + r_x)$. In the limit when the width $r_x$ approaches zero, the partial distributions $P(k|x)$ converge to the degree densities $f(k,x)$. The procedure has been carried out for different $x_0$ values ranging in the $[0.5, 1]$ interval with a width $r_x = 0.01$, small enough to obtain an accurate approximation to the degree densities. Due to symmetry considerations the $[0, 0.5)$ interval has not been scrutinized. 

Fig. \ref{fig:4} shows the stationary degree densities $f(k,x)$ obtained from Eq. \ref{eq:7} for the threshold model, next to the partial degree distributions $P(k|x)$ obtained by numerical simulation, when $\epsilon = 0$. Each curve depicted in Fig. \ref{fig:4} col. (2) measures the probability for finding a node with with connectivity degree $k$ and state $x$ satisfying the inequality $|x - x_0| < r_x$ in a network of the ensemble. It can be noted that the partial distributions exhibit a power-law scaling with different exponents that depend on the state that characterizes the subset of nodes. The scaling exponent is highest ($\gamma \simeq 3.9$) for the subset of nodes with states around the most peripheric sample state, $x_0 = 0.99$. The exponent decreases as the sample state becomes more moderate, and is lowest ($\gamma \simeq 2.4$) for the subset of nodes with states around the central state $x = 0.5$. The numerical results exhibit a good agreement with the analytical predictions in Fig. \ref{fig:4} col. (1) and provide evidence of the multiscaling phenomenon observed in the stationary degree densities.

\section{Conclusions}

In summary, in this paper we have studied by numerical simulation the scaling phenomena exhibited by the connectivity degrees in the threshold model. We have shown that the power-law scaling in the degree distribution is a robust trait in the presence of heterogeneity, however significant details such as the scaling exponent or the presence of hubs in the network depend on the form of such heterogeneity in the attachment mechanism. We have also shown that the results of the numerical simulations exhibit a fair agreement with the stationary distributions of heterogeneous preferential attachment networks analytically obtained in the thermodynamic limit. The existence of finite-size effects introduces slight discrepancies in the numerical results that only increase as the threshold value becomes very small.

The robustness evidenced by the scaling phenomenon in the degree distribution points out that heterogeneity provides a more general mechanism for the emergence of power-law distributions, with regards to the original preferential attachment mechanism suggested by Barabási and Albert. The values of the scaling exponents ($2.2 - 2.9$ in the numerical simulations, $2.5 - 3.0$ in the analytical calculations) also fit nicely within the range of values empirically observed in real networks. We suggest that this fact may help explain the ubiquity of real scale-free networks that otherwise differ in other topological aspects. The presence of heterogeneity thus may constitute a source of differentiation behind the behavior exhibited by less universal metrics.

Finally, we have shown that the presence of heterogeneity induces a multiscaling phenomenon in the degree densities of preferential attachment networks. Such multiscaling is not found in their homogeneous counterparts and hence represents a signature of heterogeneity in the topology of preferential attachment networks. We have shown that this phenomenon can be experimentally observed by computing the partial degree distributions on subsets of network nodes with states in small neighborhoods of the state space. This result provides an experimental procedure to check the presence of heterogeneity in the evolution of real networks when this process is suspected of having arisen from some form of preferential attachment.

\section*{Acknowledgments}
\label{sec:Acknowledgment}
This work has been supported by the Spanish Ministry of Education and Science under Project 'Ingenio Mathematica (i-MATH)' No.~CSD2006-00032 and Project No.~MTM2006-15533, and GESAN, S.A. under Project No.~SE05-0230-01.

% Referencias


\begin{thebibliography}{99}

\bibitem{str01} S. H. Strogatz, ``Exploring complex networks''. \emph{Nature} \textbf{410} (March 2001) : 268-276.
\bibitem{wuc03} S. Wuchty, E. Ravasz and A.-L. Barabási, ``The architecture of biological networks''. In T.S. Deisboeck, J. Y. Kresh and T.B. Kepler (eds.), \emph{Complex Systems in Biomedicine}. New York: Kluwer Academic, 2003.
\bibitem{wu03} F. Wu, B. A. Huberman, L. A. Adamic and J. Tyler, ``Information flow in social groups''. \emph{Physica A} \textbf{337} (2004) : 327-335.
\bibitem{yoo02} S.-H. Yook, H. Jeong and A.-L. Barabási, ``Modeling the Internet's large-scale topology''. \emph{Proc. Natl. Acad. Sci. USA} \textbf{99} (2002) : 13382-13386.
\bibitem{fer01a} R. Ferrer i Cancho and R. V. Solé, ``The small world of human language''. \emph{Proc. Roy. Soc. Lond. B - Biol. Sci.} \textbf{268} (2001) : 2261-2265.
\bibitem{alb02} R. Albert and A.-L. Barabási, ``Statistical mechanics of complex networks''. \emph{Rev. Mod. Phys.} \textbf{74 No. 1} (2002) : 47-97.
\bibitem{new03a} M. E. J. Newman, ``The structure and function of complex networks''. \emph{SIAM Review} \textbf{45} (2003) : 167-256.
\bibitem{dor02a} S. N. Dorogovtsev and J. F. F. Mendes, ``Evolution of networks''. \emph{Adv. Phys.} \textbf{51} (2002) : 1079-1187.
\bibitem{sim55} H. A. Simon, ``On a class of skew distribution functions''. \emph{Biometrika} \textbf{42} (1955) : 425-440.
\bibitem{pri65} D. J. S. Price, ``Networks of scientific papers''. \emph{Science} \textbf{149} (1965) : 510-515.
\bibitem{pri76} D. J. S. Price, ``A general theory of bibliometric and other cumulative advantage processes''. \emph{J. Amer. Soc. Inform. Sci.} \textbf{27} (1976) : 292-306.
\bibitem{bar99a} A.-L. Barabási and R. Albert, ``Emergence of scaling in random networks''. \emph{Science} \textbf{286} (1999) : 509-512.
\bibitem{bar03a} A.-L. Barabási and E. Bonabeau, ``Scale-free networks''. \emph{Sci. Am.} \textbf{288} (2003) : 60-69.
\bibitem{bar99b} A.-L. Barabási, R. Albert and H. Jeong, ``Mean-field theory for scale-free random networks''. \emph{Physica A} \textbf{272} (1999) : 173-197.
\bibitem{dor00b} S. N. Dorogovtsev, J. F. F. Mendes and A. N. Samukhin, ``Structure of growing networks with preferential linking''. \emph{Phys. Rev. Lett.} \textbf{85} (2000) : 4633-4636.
\bibitem{kra01} P. L. Krapivsky and S. Redner, ``Organization of growing random networks''. \emph{Phys. Rev. E} \textbf{63} (2001) : 066123.
\bibitem{alb00} R. Albert and A.-L. Barabási, ``Topology of evolving networks: local events and universality''. \emph{Phys. Rev. Lett.} \textbf{85} (2000) : 5234-5237.
\bibitem{dor00} S. N. Dorogovtsev and J. E. F. Mendes, ``Evolution of networks with aging of sites''. \emph{Phys. Rev. E} \textbf{62} (2000) : 1842-1845.
\bibitem{erg02} G. Ergün and G. J. Rodgers, ``Growing random networks with fitness''. \emph{Physica A} \textbf{303} (2002) : 261-272.
\bibitem{ama04} L. A. N. Amaral and J. M. Ottino, ``Complex networks: Augmenting the framework for the study of complex systems''. \emph{Eur. Phys. J. B} \textbf{38} (2004) : 147-162.
\bibitem{yoo01} S.-H. Yook, H. Jeong, A.-L. Barabási and Y. Tu, ``Weighted evolving networks''. \emph{Phys. Rev. Lett.} \textbf{86} (2001) : 5835-5838.
\bibitem{barr04} A. Barrat, M. Barthelemy and A. Vespignani, ``Weighted evolving networks: coupling topology and weights dynamics''. \emph{Phys. Rev. Lett.} \textbf{92} (2004) : 228701.
\bibitem{bart04} M. Barthelemy, A. Barrat, R. Pastor-Satorras and A. Vespignani, ``Characterization and modelling of weighted networks''. Proceedings, \emph{Complex networks: structure, function and processes}. \emph{Phys. A} \textbf{346} (2005) : 34.
\bibitem{bog04} M. Boguña, R. Pastor-Satorras, A. Díaz-Guilera and A. Arenas, ``Models of social networks based on social distance attachment''. \emph{Phys. Rev. E} \textbf{70} (2004) : 056122.
\bibitem{gra06} A. Grabowski and R. A. Kosinski, ``Evolution of a social network: The role of cultural diversity''. \emph{Phys. Rev. E} \textbf{73} (2006) : 016135.
\bibitem{bia01} G. Bianconi and A.-L. Barabási, ``Competition and multiscaling in evolving networks''. \emph{Europhys. Lett.} \textbf{54} (2001) : 436.
\bibitem{gom04} J. Gómez-Cardeñes and Y. Moreno, ``Local versus global knowledge in the Barabási-Albert scale-free network model''. \emph{Phys. Rev. E} \textbf{69} (2004) : 037103.
\bibitem{soa05} D. J. B. Soares, C. Tsallis, A. M. Mariz and L. R. da Silva, ``Preferential attachment growth model and nonextensive statistical mechanics''. \emph{Europhys. Lett.} \textbf{70} (2005) : 70-76.
\bibitem{thu05} S. Thurner and C. Tsallis, ``Nonextensive aspects of self-organized scale-free gas-like networks''. \emph{Europhys. Lett.} \textbf{72} (2005) : 197-203.
\bibitem{alo06} R. Alonso-Sanz, ``A structurally dynamic cellular automaton with memory''. \emph{Chaos, Solitons and Fractals} \textbf{32} (2006) : 1285-1295.
\bibitem{san06} A. Santiago and R. M. Benito, ``An extended formalism for preferential attachment in heterogeneous complex networks''. Subm. to \emph{Phys. A} (2006).
\end{thebibliography}
\end{document}